\documentclass[twocolumn,prb,showpacs,groupedaddress,preprintnumbers,amsmath,amssymb]{revtex4}
\usepackage{graphicx}
\usepackage{dcolumn}

\begin{document}
\title{Structure factor and thermodynamics of rigid dendrimers in solution}

\author{L. Harnau$^{a}$, S. Rosenfeldt$^b$, and M. Ballauff$^b$}
\affiliation{$^a$Max-Planck-Institut f\"ur Metallforschung,  
Heisenbergstr.\ 3, D-70569 Stuttgart, Germany, 
\\and Institut f\"ur Theoretische und Angewandte Physik, 
Universit\"at Stuttgart, Pfaffenwaldring 57, D-70569 Stuttgart, Germany\\
$^b$Physikalische Chemie I, University of Bayreuth, 
D-95440 Bayreuth, Germany}

\date{\today}

\begin{abstract}
The ''polymer reference interaction site model'' (PRISM) integral equation 
theory is used to determine the structure factor of rigid dendrimers in 
solution. The theory is quite successful in reproducing experimental structure 
factors for various dendrimer concentrations. In addition, the structure factor 
at vanishing scattering vector is calculated via the compressibility equation 
using scaled particle theory and fundamental measure theory. The results as 
predicted by both theories are systematically smaller than the experimental 
and PRISM data for platelike dendrimers.
\end{abstract}
\maketitle

\section{Introduction}
Dendrimers are synthetic macromolecules with defined architecture 
that are synthesized by iterative controlled reaction steps. 
Starting from a focal unit, subsequent generations are 
connected to this initial core which results in a tree-like structure 
\cite{Voe:99,tcc:01,ball:04,liko:05}. Today it is well-understood that 
dendrimers composed of flexible units adopt a so-called dense core 
structure, that is, the average segmental density has its maximum at 
the center of the molecule \cite{tcc:01,ball:04,liko:05}. This is 
easily derived from the fact that flexible dendrimers can assume a 
great number of conformations in which the terminal groups can fold 
back. Hence these flexible dendrimers do not exhibit a well-defined 
surface given by the terminal groups of the last generation. The average 
density profile thus derived can be used to calculate the interaction 
of flexible dendrimers in solution.  Thus these structure are 
well-understood by now \cite{holger:jcp:03}.

\begin{figure}
\includegraphics[width=7cm,clip]{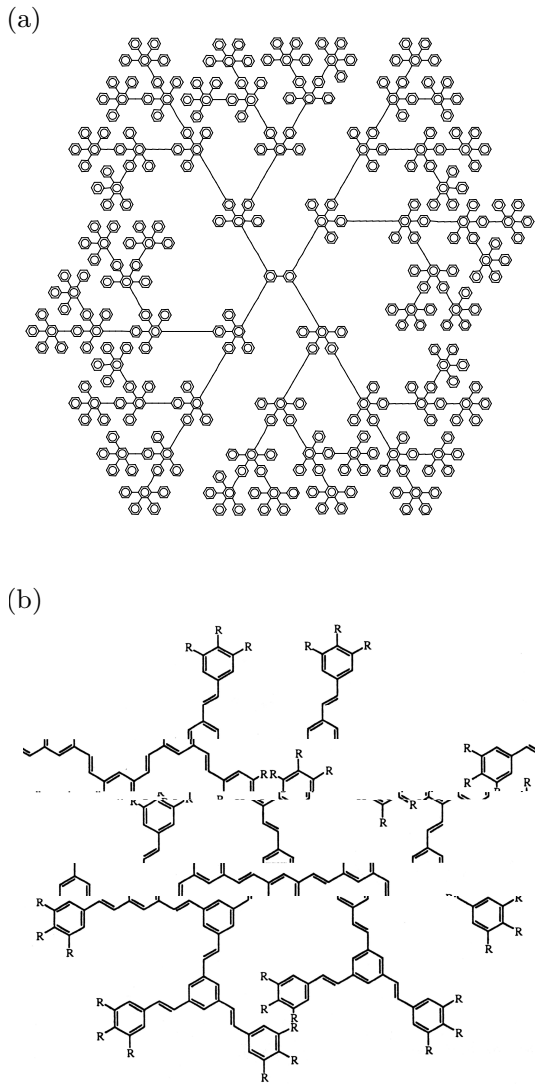}
\caption{(a) Chemical structure of the polyphenylene dendrimer of the fourth generation.
(b) Chemical structure of the stilbenoid dendrimer of the third generation
\mbox{(R $ = OC_6H_{13}$).}}
\label{fig0}
\end{figure}

Much less attention has been paid to dendrimers consisting of rigid units \cite{muel:99,wind:02,rose:04,rose:05,carb:06,meie:98,meie:00,meie:03,rose:06}. 
Figure \ref{fig0} (a) shows the chemical structure of such a rigid dendrimer which 
is solely composed of stiff units \cite{muel:99,wind:02,rose:05,carb:06}.
While the dendritic scaffold of flexible dendrimers in solution can adopt a 
large number of conformations which follow from rotations about various bonds,
dendrimers consisting of such rigid units exhibit a rather well defined structure 
in solution. This fact has been shown recently by small-angle neutron scattering 
(SANS) \cite{rose:04,rose:05,rose:06}. Hence, these systems may serve as model 
systems for interacting {\it monodisperse} particles in statistical physics. 

Here we present a comprehensive discussion of the interaction of rigid dendrimers 
in solution. Two systems involving rigid dendrimers are examined in the present 
paper, namely polyphenylene dendrimers of the fourth generation in solution 
(see Fig.~\ref{fig0} (a)) and stilbenoid dendrimers of the third generation 
in solution (see Fig.~\ref{fig0} (b)). The problem to be addressed is that of 
structural properties over a range of dendrimer concentrations. In general, the 
effect of mutual interaction of dissolved species with number density $\rho=N/V$ in a 
scattering experiment can be embodied in the structure factor $S(q,\rho)$ defined as
\begin{equation}
I(q,\rho) = \frac{N}{V}I_0(q)S(q,\rho) = \frac{N}{V}V_p^2(\Delta \bar\rho)^2P(q)S(q,\rho)
\label{equ_S(q)}
\end{equation}
where $I_0(q)$ is the scattering intensity of the single particle, $V_p$ the volume 
of the particle, $P(q)$ the form factor (normalized to unity for $q=0$) and $q$ is 
the magnitude of the scattering vector ($q = (4\pi/\lambda)sin(\theta/2)$, $\lambda$: 
wavelength of radiation, $\theta$: scattering angle). $\Delta \bar\rho$ 
is the contrast of the solute resulting from the difference of the average scattering 
length density and the scattering length density of the solvent. 
Up to now, mutual interaction has 
only been addressed because scattering experiments require finite concentrations 
in order to obtain data with a reasonable statistics \cite{rose:04,rose:05,rose:06}. 
Hence, the effect of mutual interaction of the dissolved dendrimers has only been 
considered when extrapolating the data to vanishing concentrations. Details of 
such a data evaluation may be found in Ref.~\cite{rose:05}. Here we determine 
the  correlation functions and structure factors $S(q,\rho)$ within the framework of an 
interaction site model and compare these results to experimental data 
\cite{rose:04,rose:05,rose:06}. Moreover, the validity of the predictions of both 
scaled particle theory and fundamental measure theory are investigated. A systematic 
comparison is made between the results of the theories and experimental data 
for two differently shaped dendrimers in solution, in order to quantify the 
influence of the shape of the particles on the quality of the predictions of the 
theories. We demonstrate 
that rigid dendrimers present a new class of strictly monodisperse model colloids 
whose interaction can be conveniently studied in solution. In this way we show that 
these systems allow us to re-consider the basic statistical mechanics of rigid objects 
in solution and to compare theory and experiment for the first time in a rigorous fashion.

The paper is organized as follows: In the next section we briefly recapitulate 
the analytical models necessary to calculate the scattering intensity for a system 
of interacting rigid bodies. Thereafter the theory will be compared to experimental 
results obtained for the rigid dendrimers shown in Figs.~\ref{fig0} (a) and (b). 
In section IV these results are compared to the fundamentals of the thermodynamics 
of rigid bodies. Special attention is paid to a discussion of exact virial coefficients.
A brief conclusion will summarize the results at the end.

\section{RISM and PRISM theories}
The systems under investigation are solutions. However, in view of the mesoscopic 
scale of the particles, the solvent will be considered as a structureless continuum.
Spatial pair correlations of an isotropic fluid of identical particles, each 
carrying $n$ distinct interaction sites, are characterized by a set of 
intermolecular site-site total correlation functions $h_{ij}(r,\rho)$, where the 
indices $i$ and $j$ run over sites on each of two particles and $\rho$ is the 
particle number density. These functions are related to a set of intermolecular 
site-site direct correlation functions  $c_{ij}(r,\rho)$ by the generalized 
Ornstein-Zernike relations of the ''reference interaction site model'' (RISM), 
which in Fourier space read \cite{chan:72,chan:82}
\begin{eqnarray} \label{eq1}
h_{ij}(q,\rho)&=&\sum\limits_{m, o =1}^n\omega_{im}(q,\rho)c_{mo}(q,\rho)\nonumber
\left(\omega_{oj}(q,\rho)+\rho h_{oj}(q,\rho)\right)\,,
\end{eqnarray}
where the $\omega_{ij}(q,\rho)$ are the Fourier transforms of the intramolecular 
correlation functions. The set of generalized Ornstein-Zernike equations must be 
supplemented by a set of closure relations. If the interaction sites are simply 
the centers of exclusion spheres of diameter $d$, to account for steric effects, 
a convenient closure is the Percus-Yevick approximation \cite{chan:72,hans:86}
\begin{eqnarray} \label{eq2}
h_{ij}(r,\rho)=-1\,,\,\,r\le d\,,\hspace*{0.5cm}
c_{ij}(r,\rho)=0\,,\,\,r> d\,.
\end{eqnarray}
The experimentally accessible structure factor $S(q,\rho)$ is defined as
\begin{eqnarray} \label{eq3}
S(q,\rho)=1+\rho\frac{h(q,\rho)}{P(q,\rho)}\,,
\end{eqnarray}
where 
\begin{eqnarray} \label{eq4}
h(q,\rho)=\frac{1}{n^2}\sum\limits_{m, o =1}^nh_{mo}(q,\rho)
\end{eqnarray}
is the particle-averaged total correlation function. The particle-averaged 
intramolecular correlation function
\begin{eqnarray} \label{eq5}
P(q,\rho)=\frac{1}{n^2}\sum\limits_{m, o =1}^n\omega_{mo}(q,\rho)
\end{eqnarray}
characterizes the geometry of the distribution of the sites, 
and hence the geometric shape of the particles. While the particle-averaged 
intramolecular correlation function accounts for the interference of radiation 
scattered from different parts of the same particle in a
scattering experiment, the local order in the fluid is 
characterized by $h(q,\rho)$ or $S(q,\rho)$. 

The RISM has been proved to be a successful theory of the pair 
structure of many molecular fluids (for a review see Ref.~\cite{mons:90}).
In the case of macromolecular and colloidal systems, with very large 
numbers of interaction sites, the number of coupled RISM equations becomes 
intractable, and a considerable simplification follows from the assumption 
that the direct correlation functions $c_{ij}(q,\rho)$ are independent of the 
indices $i$ and $j$. This leads to the ''polymer reference interaction site model'' 
(PRISM) theory first applied by Schweizer and Curro to long flexible polymers 
\cite{schw:87}. PRISM neglects end effects in that case. The resulting single 
generalized Ornstein-Zernike equation of the PRISM reads 
\begin{eqnarray} \label{eq8}
h(q,\rho)=c(q,\rho)P^2(q,\rho)+\rho c(q,\rho)h(q,\rho)P(q,\rho)\,,
\end{eqnarray}
where $c(q,\rho)=\sum_{m,o=1}^nc_{mo}(q,\rho)$. The PRISM integral equation theory has 
been successfully applied to various systems, such as rodlike viruses 
\cite{yeth:96,yeth:97,yeth:98,harn:00}, platelike colloids \cite{harn:01,li:05,webe:07} 
and dendrimers \cite{rose:06}, polymers \cite{schw:97,harn:01a}, mixtures of 
spherical colloids and semiflexible polymers \cite{harn:02}, and bottlebrush 
polymers \cite{boli:07}. Moreover, it has been demonstrated recently that the
simpler PRISM theory yields results in good agreement with the more elaborate 
RISM calculations for lamellar colloids \cite{cost:05}.

\begin{figure}
\includegraphics[width=6.2cm,clip]{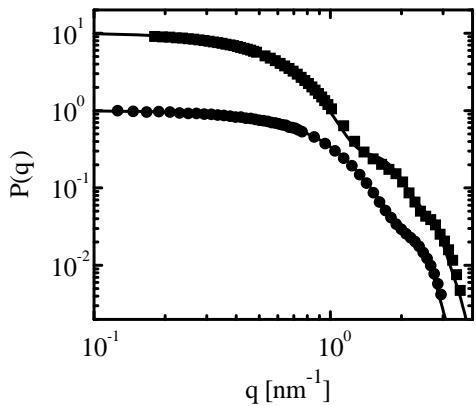}
\caption{The form factor $P(q)=P(q,\rho \to 0)$ of polyphenylene dendrimers 
of the fourth generation  \cite{rose:05} (squares and upper line) and stilbenoid 
dendrimers of the third generation \cite{rose:06} (circles and lower line) as 
obtained by small angle neutron scattering (symbols). The lines represent the 
calculated form factors. For reasons of clarity, the upper data set and line have 
been shifted up.}
\label{fig1}
\end{figure}

For flexible polymers or dendrimers the particle-averaged intramolecular 
correlation function $P(q,\rho)$ depends on the particle number density and 
follows from a statistcal average over particle configurations. In the limit
$\rho\to 0$ the particle-averaged intramolecular correlation function 
reduces to the form factor \mbox{$P(q)\equiv P(q,\rho\to 0)$}. In the 
case of rigid particles $P(q,\rho)$ is independent of the particle number density
because the particles are not deformed due to intermolecular interactions 
for typical concentrations in the fluid state.

\section{Rigid polyphenylene and stilbenoid dendrimers in solution}
\subsection{Form factors}
It has been demonstrated that the building units of the dendritic scaffold 
of polyphenylene dendrimers of the fourth generation (see Fig.~\ref{fig0} (a)) 
are rather well-localized and no back folding of the terminal groups occurs 
\cite{rose:04,rose:05}. A general feature of the polyphenylene 
building block is its finite angle formed by two subsequent phenyl groups
\cite{wind:02,carb:06}. This is due to the strong repulsion between their 
$o$-hydrogen atoms that prevent a co-planar conformation. As a consequence, 
the polyphenylene dendrimers exhibit a three-dimensional structure which 
has been found by SANS \cite{rose:04,rose:05}. 

Such steric hindrance does 
not exists in dendritic scaffolds set up from stilbenoid units 
\cite{meie:98,meie:00,meie:03}. Starting from a central phenyl group all
subsequent generations are built up by {\it trans}-stilben units, only the 
terminal groups are substituted by hexyloxy groups in order to ensure 
better solubility in common solvents such as toluene (see Fig.~\ref{fig0} (b)). 
Full conjugation 
in {\it trans}-stilbene can be achieved in the completely planar conformation. 
However, the potential energy for a slight torsion around the single bonds 
is low in the ground state of trans-stilbene. Both molecular modeling
\cite{meie:00} and a small-angle scattering study \cite{rose:06} have 
demonstrated that a stilbenoid dendrimer of the third generation exhibits 
a relatively compact platelike structure.

The measured and calculated form factors $P(q)$ of both polyphenylene 
dendrimers of the fourth generation and stilbenoid dendrimers of the third 
generation are shown in Fig.~\ref{fig1}. The form factor of the stilbenoid 
dendrimers agrees with the one calculated numerically for a circular platelet 
with a radius \mbox{$R=2.4$ nm} and thickness \mbox{$L=1.8$ nm} according 
to
\begin{eqnarray} \label{eq14}
P(q)&=&4\int\limits_0^1 dx\,\frac{J^2_1(qR\sqrt{1-x^2})}{(qR\sqrt{1-x^2})^2}
\frac{\sin^2(qLx/2)}{(qLx/2)^2}\,,
\end{eqnarray}
where $J_1(x)$ denotes the cylindrical Bessel function of first-order 
\cite{higg:94}. Equation (\ref{eq14}) follows from  Eq.~(\ref{eq5}) 
by replacing the double sums by double integrals corresponding to a 
continuous distribution of interaction sites ($n\to\infty$) and by using 
$\omega_{mo}(q,\rho)=\sin(ql_{mo})/(ql_{mo})$, where $l_{mo}\to 0$
is the bond length between sites $m$ and $o$ on the same cylindrical particle.

\begin{figure}
\includegraphics[width=6.2cm,clip]{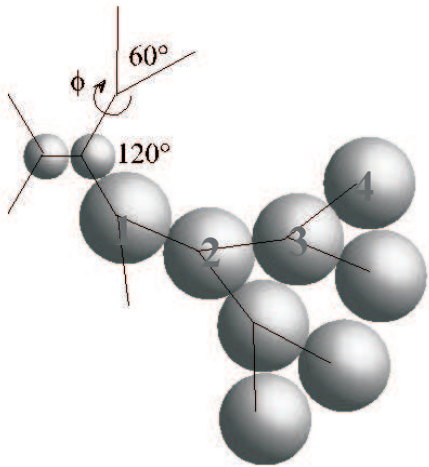}
\caption{Modeling of the the form factor $P(q)$ of the polyphenylene dendrimer of the 
fourth generation. The dendritic scaffold is set up of a central unit consisting 
of two small spheres (distance: 0.4 nm) describing the central biphenyl unit. The four 
dendrons were modeled by 8 spheres (distance: 0.87 nm), respectively,  having a diameter 
of 0.8 nm. Both parameters were used as a fitting parameter. For the sake of clarity 
the size of the spheres has been depicted slightly smaller. The angles between the 
different units is indicated in the graph. The torsional angle was chosen at random 
for each dendron.}
\label{fig2}
\end{figure}

The polyphenylene dendrimers were modelled by a central unit consisting of 
two small spheres (diameter: \mbox{$0.4$ nm}). This describes the central 
biphenyl unit (see Fig.~\ref{fig2}). The four dendrons were mimicked by 
eight spheres (diameter: \mbox{$0.8$ nm}) with a center to center distance 
of \mbox{$0.87$ nm}. This distance is the approximate length of the two 
phenyl groups that connect two subsequent shells. Each sphere comprises 
five benzene rings that constitute a branching group. The endgroups are 
made of an equal number of phenyl rings. The form factor of the polyphenylene 
dendrimers has been modelled by taking an average over approximately 
500 conformers generated by randomly choosing the torsion angles for 
each dendron (see Ref.~\cite{rose:05} for further details). The good agreement 
between the experimental and theoretical results (Fig.~\ref{fig1}) 
demonstrates that a full understanding of the spatial structure of both
polyphenylene dendrimers of the fourth generation and stilbenoid dendrimers 
of the third generation has been achieved \cite{rose:05,rose:06}. The ability 
to construct coarse-grained models in such a way as to reproduce experimentally 
determined form factors is important as it allows one to quickly study 
within PRISM theory various material systems based on input of a small amount 
of intramolecular information. 

\subsection{Structure factors}
\begin{figure}
\includegraphics[width=6.2cm,clip]{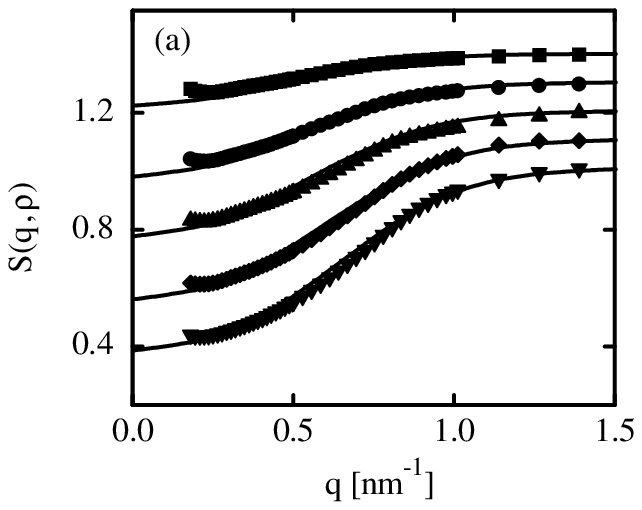}\\
\vspace*{0.5cm}
\includegraphics[width=6.2cm,clip]{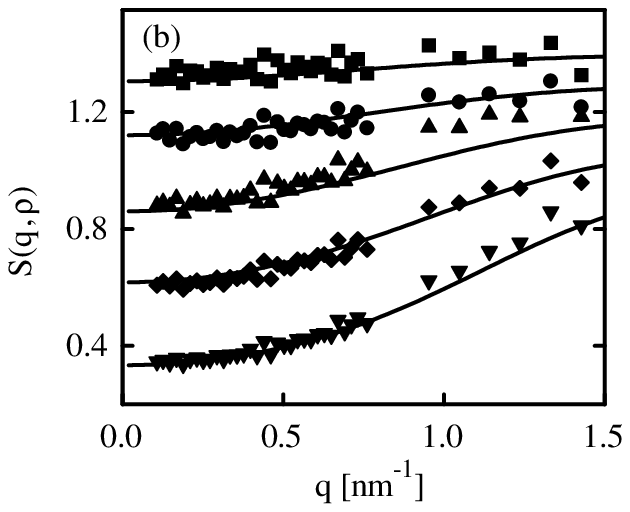}
\caption{Experimentally determined structure factors for polyphenylene dendrimers 
of the fourth generation \cite{rose:05} in (a) and  stilbenoid dendrimers of the 
third generation \cite{rose:06} in (b) together with the results of the theoretical 
predictions of the PRISM integral equation theory (Eqs.~(\ref{eq3}) and (\ref{eq8})). 
The volume fraction of the dendrimers increases from top to bottom: 
$\phi=0.016, 0.032, 0.046, 0.065, 0.08$ in (a);
$\phi=0.009, 0.019, 0.039, 0.061, 0.1$ in (b). For reasons of clarity, the upper data 
sets and lines have been shifted up by 0.1, 0.2, 0.3, 0.4, respectively.}
\label{fig3}
\end{figure}

In Figs.~\ref{fig3} (a) and (b) the experimental structure factors $S(q,\phi)$ 
are compared to the results of the integral equation theory for the PRISM. We 
have used the calculated form factors $P(q)$ (see the solid lines in Fig.~\ref{fig1}) 
as input into the generalized Ornstein-Zernike equation, i.e., $P(q,\phi)=P(q)$ 
in Eqs.~(\ref{eq3}) and (\ref{eq8}). The particle number density is given by
$\rho=\phi/V_p$, where $\phi$ is the volume fraction and $V_p$ is the volume 
of an individual particle. The generalized Ornstein-Zernike equation is 
solved numerically together with the Percus-Yevick closure. From Fig.~\ref{fig3} 
it is apparent that the PRISM integral equation theory is rather accurate. The 
magnitude and the scattering vector range of the suppression of $S(q,\rho)$, i.e., 
the deviations from the value 1 at small scattering vectors, are characteristic 
for the size and the shape of the dendrimers as well as the volume fraction. 
The experimentally observed small upturns of $S(q,\rho)$ at low scattering vectors 
$q$ for polyphenylene dendrimers in solution (Fig.~\ref{fig3} (a)) indicate
the presence of a small amount of aggregates due to attractive interactions
which are not taken into account in the theoretical calculations.
On the basis of our experience with both PRISM and RISM we expect that the results 
of the integral equation theory for the RISM would lead to very similar results, 
provided the same form factors are used \cite{cost:05}. 

\section{Thermodynamic properties}
The structure factor provides a direct link with thermodynamics via the compressibility 
equation \cite{hans:86}
\begin{eqnarray} \label{eq6}
\lim\limits_{q\to 0}S(q,\rho)
=\rho k_BT\kappa_T(\rho)\,,
\end{eqnarray}
where $\kappa_T(\rho)$ is the isothermal compressibility. The osmotic pressure 
$P(\rho)$ (equation of state) then follows from
\begin{eqnarray} \label{eq7}
\frac{P(\rho)}{k_BT}=\int\limits_0^\rho d\rho'\,S^{-1}(q=0,\rho')\,.
\end{eqnarray}
Various attempts have been made to develop accurate theories for the equation of state of 
fluids consisting of non-spherical particles:

(a) Scaled particle theory \cite{reis:59,reis:60,lebo:65}, which is very successful 
for hard sphere fluids, has been extended to prolate and oblate ellipsoids of 
revolution \cite{cott:79}, however with moderate success when gauged against Monte
Carlo simulations \cite{muld:85}. Recently, it has been shown \cite{over:05} that 
the results of scaled particle theory \cite{savi:81,boub:75,over:05} for platelike 
particles or the closely related model of hard cut spheres are in disagreement 
with computer simulation data 
\cite{fren:82,eppe:84,veer:92,bate:99,zhan:02a,zhan:02b,beek:04}.

(b) Onsager theory \cite{onsa:42,onsa:49}, based on the second viral coefficient 
alone, can be ''rescaled'' \cite{pars:79,lee:87,lee:89}. Also this semi-empirical 
procedure leads to reasonably good results for rodlike particles, it is much less 
satisfactory for platelike particles \cite{wens:04}.

(c) Many theoretical studies on hard sphere fluids and depletion agents use the 
so-called free volume theory \cite{lekk:92}, in which the free volume accessible 
to a single particle plays a major role. Recently, this free volume theory has 
been studied within a fundamental measure theory \cite{over:05}. However, 
it has been demonstrated that the resulting third virial coefficients of the 
equation of state for both hard cylinders and hard cut spheres differ from 
computer simulation results (see tables I and II in Ref.~\cite{over:05}). 
Theoretical approaches based on fundamental measure theory do not yield correct 
third virial coefficients and equation of states due to the occurrence 
of so-called ''lost cases'', i.e., the fact configurations of three particles 
with pairwise overlap but no triple overlap do not contribute to thermodynamic 
properties (see, e.g., Refs.~\cite{rose:88,tara:97}).

These earlier theoretical and computer simulation studies demonstrate that the 
understanding of thermodynamic properties of non-spherical particles needs to 
be improved. Here we model the measured inverse structure factor $S^{-1}(q=0,\rho)$ 
extrapolated to vanishing scattering vectors of stilbenoid dendrimers of the third 
generation in terms of the so-called $y3$-expansion \cite{barb:79,barb:80}
\begin{eqnarray} 
S^{-1}(q=0,\rho)&=&\frac{1+2(B_2-2)\phi+(3B_3-8B_2+6)\phi^2}{(1-\phi)^4}
\nonumber\\
\label{eq9}
\\&\approx&1+2B_2\phi+3B_3\phi^2+O(\phi^3)\,.\label{eq10}
\end{eqnarray}
The $y3$ theory reproduces the exact second and third virial coefficients, 
$B_2$ and $B_3$, respectively. However, its practical applicability is limited 
due to the difficult numerical evaluation of the third virial coefficient
$B_3$ in the case of non-spherical particles \cite{harn:02b}, while the second 
virial coefficient $B_2$ for an isotropic hard convex body fluid is known exactly 
(see Ref.~\cite{harn:06} and references therein):
\begin{eqnarray} \label{eq11}
B_2=\frac{1}{V_p}\left(V_p+A_p\tilde{R_p}\right)\,.
\end{eqnarray}
Here $A_p$ and ${\tilde R_p}=(1/4\pi)\int dA_p\,H_p$ are the surface area and the 
mean radius, respectively, where the local mean curvature is denoted as $H_p$. For a 
circular platelet of radius $R$ and thickness $L$ one has $V_p=\pi R^2L$, 
$A_p=2\pi R(R+L)$, and ${\tilde R}_p=\pi R/4+L/4$. For platelike stilbenoid dendrimers 
of the third generation the second virial coefficient $B_2=5.54$ as calculated from 
Eq.~(\ref{eq11}) with \mbox{$R=2.4$ nm} and \mbox{$L=1.8$ nm} agrees with the 
experimentally determined second virial coefficient. 

\begin{figure}
\includegraphics[width=6.2cm,clip]{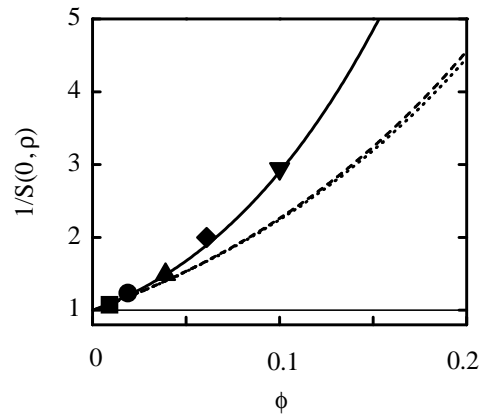}
\caption{Inverse structure factor $S^{-1}(q=0,\rho)$ extrapolated to vanishing 
scattering vectors of stilbenoid dendrimers of the third generation \cite{rose:06}
(with the same symbol code as in Fig.~\ref{fig3} (b)). The dashed line follows from 
the scaled particle theory according to Eqs.~(\ref{eq9}), (\ref{eq11}), and (\ref{eq12}) 
with $B_3^{(SPT)}=16.952$ while the dotted line represents the results of the 
fundamental measure theory as obtained from Eqs.~(\ref{eq9}), (\ref{eq11}), and 
(\ref{eq13}) with $B_3^{(FMT)}=16.432$. The solid line shows the results 
as obtained from  Eqs.~(\ref{eq9}) and (\ref{eq11}) with $B_3=34.129$.}
\label{fig4}
\end{figure}

In the framework of both scaled particle theory and Rosenfeld's fundamental 
measure theory $S^{-1}(q=0,\rho)$ is also given by Eqs.~(\ref{eq9}) - (\ref{eq11}) 
but the third virial coefficient is given by 
\begin{eqnarray} \label{eq12}
B_3^{(SPT)}=\frac{1}{V_p^2}
\left(V_p^2+2{\tilde R}_pA_pV_p+\frac{1}{3}{\tilde R}_p^2A_p^2\right)
\end{eqnarray}
within scaled particle theory and
\begin{eqnarray} \label{eq13}
B_3^{(FMT)}\frac{1}{V_p^2}
\left(V_p^2+2{\tilde R}_pA_pV_p+\frac{1}{12\pi}A_p^3\right)
\end{eqnarray}
within fundamental measure theory. In Fig.~\ref{fig4} the experimentally 
determined inverse structure factor \mbox{$S^{-1}(q=0,\rho)$} extrapolated to vanishing 
scattering vectors of stilbenoid dendrimers of the third generation 
is compared with the results of scaled particle theory according 
to Eqs.~(\ref{eq9}), (\ref{eq11}), and (\ref{eq12}) and fundamental measure 
theory according to Eqs.~(\ref{eq9}), (\ref{eq11}), and (\ref{eq13}). With 
increasing volume fraction the theoretical results of both 
scaled particle theory (dashed line) and fundamental measure theory (dotted line)
deviate from the experimental data (symbols). These deviations are mainly due to 
the fact that the predicted third virial coefficients $B_3^{(SPT)}=16.952$ and 
$B_3^{(FMT)}=16.432$ are too small. The results for $S^{-1}(q=0,\rho)$ 
as obtained from  Eqs.~(\ref{eq9}) and (\ref{eq11}) with $B_3=34.129$ as input
are in agreement with the experimental data (see the solid line in Fig.~\ref{fig4}).
We emphasize that the PRISM theory discussed above leads to a similar agreement 
with the experimental data as is apparent from Fig.~\ref{fig3} (b). 

Our first comparison of experimentally determined $S^{-1}(q=0,\rho)$ of platelike 
particles with the predictions of the well-known scaled particle and fundamental 
measure theory confirms earlier caveats concerning the applicability of these 
theories to freely rotating non-spherical particles.

\section{Conclusion}
We have presented a systematic application of the PRISM integral equation theory,
scaled particle theory, and fundamental measure theory to rigid dendrimers in 
solution. The main findings and conclusions may be summarized as follows.

PRISM theory is quite successful in reproducing experimental structure factors 
$S(q,\rho)$ of both polyphenylene dendrimers of the fourth generation   
[Fig.~\ref{fig3} (a)] and  stilbenoid dendrimers of the third generation 
[Fig.~\ref{fig3} (b)], provided the correct form factor $P(q)$ [Fig.~\ref{fig1}] is 
used as input into the generalized Ornstein-Zernike equation of the PRISM [Eq.~(\ref{eq6})]. 
These investigations encourage to pursue a study of charged dendrimers 
\cite{nisa:00,ohsh:01,ramz:02} within the framework of PRISM by using interaction 
sites which carry charges.

The inverse structure factor $S^{-1}(q\to 0,\rho)$ extrapolated to vanishing scattering
vectors as predicted by both scaled particle theory and fundamental measure theory 
is systematically smaller than the experimental data for platelike stilbenoid dendrimers 
of the third generation and the prediction of the PRISM integral equation theory
[Fig.~\ref{fig4}]. The substantial differences observed between the experimental 
data and the results of scaled particle theory and fundamental measure theory 
are mainly due the fact the both theories do not yield the correct third virial 
coefficient in the case of platelike particles. Hence there is a clear need to 
improve both scaled particle theory and fundamental measure theory for freely 
rotating non-spherical particles.

\end{document}